\documentstyle[12pt,epsfig,cite]{article}
\newcommand{\be}{\begin{equation}}
\newcommand{\ee}{\end{equation}}
\newcommand{\bea}{\begin{eqnarray}}
\newcommand{\eea}{\end{eqnarray}}

\newcommand{\la}{\langle}
\newcommand{\ra}{\rangle}

\newcommand{\p}{\partial}

\newcommand{\SL}[0]{{\rm SL}(2,\IR)}
\newcommand{\SLE}[0]{SL(2,C)/SU(2)}
\newcommand{\SLC}[0]{SL(2,C)}
\def\half{\frac{1}{2}}

\def\CZ{{\mathcal{Z}}}

\def\IB{\relax\hbox{$\inbar\kern-.3em{\rm B}$}}
\def\IC{\relax\hbox{$\inbar\kern-.3em{\rm C}$}}
\def\ID{\relax\hbox{$\inbar\kern-.3em{\rm D}$}}
\def\IE{\relax\hbox{$\inbar\kern-.3em{\rm E}$}}
\def\IF{\relax\hbox{$\inbar\kern-.3em{\rm F}$}}
\def\IG{\relax\hbox{$\inbar\kern-.3em{\rm G}$}}
\def\IGa{\relax\hbox{${\rm I}\kern-.18em\Gamma$}}
\def\IH{\relax{\rm I\kern-.18em H}}
\def\IK{\relax{\rm I\kern-.18em K}}
\def\IL{\relax{\rm I\kern-.18em L}}
\def\IP{\relax{\rm I\kern-.18em P}}
\def\IR{\relax{\rm I\kern-.18em R}}
\def\IC{\relax{\rm I\kern-.18em C}}
\def\IZ{\relax{\rm Z\kern-.5em Z}}

\begin{document}
\begin{titlepage}

\begin{flushright}
OUTP-00-50P\\
hep-th/0011056\\
\end{flushright}

\begin{center}
{\Large \bf Logarithmic operators in $\SL$ WZNW model, Singletons and $AdS_{3}/(L)CFT_{2}$ correspondence}
\vspace{0.1in}  

\vspace{0.5in}
{\large  Sanjay}
\footnote{s.sanjay1@physics.ox.ac.uk, sanjay@iopb.res.in}\\
{\small
Theoretical Physics,\\ 
University of Oxford, Oxford OX1 3NP, UK.}\\

\vspace{0.5in}

{\bfseries ABSTRACT}\\

\end{center}
\vspace{.2in}

We discuss the role of singletons and logarithmic operators in $AdS_3$ string theory in the context of $AdS_{3}/CFT_{2}$ correspondence. 

\vfill

\end{titlepage}

\newpage

\section{Introduction}
String theory in $AdS_3$ background has been
widely studied as an example of string theory in nontrivial curved
spacetime background in last one decade and recently
in the context of AdS/CFT correspondence
\cite{oldstring,newstring}. Worldsheet theory is
described by the $\SL$ WZNW model. 

Correlation functions in WZNW model satisfy Knizhnik-Zamolodchikov
(KZ) equation \cite{Knizhnik:1984nr}. In references
\cite{Kogan:1999hz,Nichols:2000mk, Hosomichi:2000bm} some solutions of
Kniznik-Zamolochikov equation for four point functions of $\SL$
primaries were found which were logarithmically singular as the two
operators approach each other. It was suggested that it might be
necessary to include contribution of logarithmic operators in operator
product expansion (OPE) and $\SL$ WZNW conformal theory is a
logarithmic conformal field theory (LCFT)\cite{Gurarie:1993xq,LCFTrefs} (see also the recent reviews \cite{rev} and references therein). The solutions involved one or more operators in non-trivial identity representation. In addition to trivial identity which is just a constant, we also have a non-trivial identity in $\SL$ representation theory (see below). This corresponds to singletons. These are special finite dimensional representations which lie at the limit of unitary bound \cite{singleton}. 

A representation of $\SL$ group \cite{Barg} element can be labeled by
eigenvalue of quadratic Casimir, $-j(j+1)$ and eigenvalue, $m$ of one
of generators say $J^3$. We have (i) Continuous Representation,
$C_{j}$, $j+\alpha\not\in \IZ$, $\alpha$ being the fractional part of
$m$ (ii) Discrete Series of lowest weight type, $D^{+}_{j}$,
$2j\not\in \IZ^{+}\cup 0$ (iii)Discrete Series of highest weight type,
$D^{-}_{j}$, $2j\not\in \IZ^{+}\cup 0$ and (iv) finite dimensional
representation $I_{j}$, $2j\in \IZ^{+}\cup 0$. We have  suppressed the
other label $m$ and fractional part of $m$. Continuous series is
unitary for $j=-\half+i\rho$, $\rho \in \IR$ and for an exceptional
interval $-\half<j<\half$, discrete series $D^{-}_{j}$ and $D^{+}_{j}$ for $j<0$ and finite dimensional representation $I_{j}$ for $j=0$ only. 

 Finite dimensional representation can be embedded in a reducible but indecomposable representation with eigenvalues of $J^{3}$ unrestricted. Singletons are part of this non-unitary indecomposable finite dimensional representation $j=0$. It should be distinguished from trivial $j=0$ representation which appears if we restrict ourselves to unitary irreducible representations. 

Singleton representation is somewhat special. Two point function of singleton modes in $AdS_{3}$ has two solutions, one of them is logarithmic (see for example \cite{Starinets:1998dt} and references therein). One can throw away the logarithmic solution by imposing vanishing flux condition at infinity in favour of the other. However it can still give logarithmic singularities in four point functions. It was shown in \cite{Kogan:1999bn} that if one considers singletons in the bulk of $AdS_{3}$, then two point function are logarithmic in boundary conformal field theory (see also \cite{Ghezelbash:1998rj}). It was conjectured \cite{Kogan:1999bn} that boundary conformal field theory is logarithmic conformal field theory. Calculation of absorption cross section for gauge bosons in the bulk of $AdS_{3}$ \cite{Kim:1998yw, Myung:1999nd} (further discussed in \cite{Lewis:1999qv} in the context of AdS/LCFT correspondence) lends some support to this conjecture.

It was argued that $j= -\half \in D^{+}_{j}$  representation
determines whether boundary CFT is logarithmic or
not\cite{Lewis:2000tn}. $j=-\half$ was considered as unitary bound in
$\SL$ WZNW model in that reference. AdS/CFT correspondence allows us
to define correlation function for $j>-\half$ beyond the bound
$j=-\half$. We show that they are well behaved unless we reach the
unitary bound $j=0$. 

At $j=0$ we have to take into account of singleton representation and correlation functions can be logarithmic. 

The paper is organized as follows. In section 2 we discuss the
origin of logarithmic operators in $AdS_3$ string theory. In section 3 it is
shown how singletons can give logarithmic correlation functions in
boundary CFT. In section 4 we discuss correlation
functions of fields with $j>-\half$ and dimension dependent
normalization of the fields in the interval $0\ge j\ge -\half$. We
conclude with brief discussion and summary.

\section{Logarithmic operators in $\SL$ WZNW model}

Primary fields of $\SL$ WZNW are labeled by the quantum numbers of global $\SL$ symmetry. Generators of global $\SL$ group satisfy. 

\bea \label{eq:SLalgeb} [J^+,J^-]=-2J^3 ~~~~
[J^3,J^{\pm}]=\pm J^{\pm} 
\eea

The quadratic Casimir is,
\bea 
C_2=\eta_{ab} J^a J^b=\half (J^+J^- + J^-J^+) -J^3J^3=-j(j+1).
\eea

A representation for the  $\SL$ generators is given by \cite{Zamolodchikov:1986bd},

\bea \label{eq:repn}
D^+=-x^2\frac{\p}{\p x}+2jx, ~~~ 
D^-=-\frac{\p}{\p x}, ~~~
D^3=x\frac{\p}{\p x}-j 
\eea

Primary fields, $\phi_j(x,z)$ satisfy,

\be
\label{ddef}
J^a\phi_j(x,z) = D^a(x) \phi_j(x,z) 
\ee

where $J^{a}$ are the zero modes of the $\SL$ current algebra and
$D^a(x)$ is as given in equation (\ref{eq:repn}). The fields $\phi_j(x,z)$ are also primary with respect to the Virasoro algebra with $L_0$ eigenvalue:

\be
\Delta_{j} = -\frac{j(j+1)}{k-2}
\ee

For the discussion of correlation function we consider Euclidean version of $\SL$, which is $\SLE$. Primary fields of the Euclidean model are given by \cite{deBoer:1998pp, Kutasov:1999xu},

\bea
\phi_j(x,z)& \sim & \left [ \left(1, -x \right) g 
\left(\begin{array}{cc}
1\\ 
-\bar{x}
\end{array}
\right)
\right] \\  \nonumber
& = & \left [(\gamma - x)(\bar{\gamma} - \bar{x})e^{2\phi} +
  e^{-2\phi}\right ]^{2j}
\eea

where $g\in\SLC$ and $\gamma, \bar{\gamma}$ and $\phi$ are $\SLC$ co-ordinates. 

The primary field $\phi_j(x,z)$ has the form of a bulk to boundary Green function for a scalar field in $AdS_3$ and $(x, \bar{x})$ has the interpretation of boundary co-ordinates\cite{deBoer:1998pp}.

Correlation functions in WZNW model satisfy a set of partial differential equations known as  Knizhnik-Zamolodchikov (KZ) equation \cite{Knizhnik:1984nr}. For two and three point functions it gives no new information. Four point functions are determined by conformal invariance up to a function, $F(x,z)$ of conformally invariant cross ratios $z=\frac{z_{12}z_{34}}{z_{13}z_{24}}, ~~~ x=\frac{x_{12}x_{34}}{x_{13}x_{24}}$. The Knizhnik-Zamolodchikov (KZ) equation for the four point functions become a partial differential equation for conformal blocks $F(x,z)$. 

Existence of logarithmic operators is signaled by the presence of
logarithmic singularities of cross ratios in four point solutions. In
reference \cite{Nichols:2000mk} we found some exact solutions with
logarithms of cross ratios in conformal blocks. 

Consider a solution of KZ equation with logarithmic singularities of
cross ratios ($j=j_1=j_3\neq -\half ~~~ j_2=j_4=0$).

\bea
\la \phi_j(x_1,z_1)\phi_0(x_2,z_2)\phi_{j}(x_3,z_3)\phi_{0}(x_4,z_4) \ra =|z_{13}|^{-2h} 
|x_{13}|^{-2j}~\biggl[ A\Bigl( \ln \left| \frac{1-x}{x} \right| +  \nonumber \\
\frac{2j+1}{k-2}\ln \left| \frac{1-z}{z} \right| \Bigr) + B  \biggr] \nonumber \\
\eea

Now taking the  $1 \rightarrow 2$ and $3 \rightarrow 4$ limit which corresponds to $x,z \rightarrow 0$ we can expand our solution as
\bea
\la \phi_j(x_1,z_1)\phi_0(x_2,z_2)\phi_{j}(x_3,z_3)\phi_{0}(x_4,z_4) \ra =|z_{13}|^{-2h}
|x_{13}|^{-2j}~\Bigl[ \ln|x_{12}|  \ln|x_{34}|   \nonumber \\
-2\ln|x_{24}| + \frac{2j+1}{k-2} \left(\ln|z_{12}| +\ln|z_{34}| - 2\ln|z_{24}| \right) +  ... \Bigr]
\eea

where we use $\ln x=\ln x_{12}+\ln x_{34} - \ln x_{13} -\ln x_{24}=\ln x_{12} + \ln x_{34} - 2 \ln x_{24} + ...$ and similarly for $z$. $'...'$ stands for subleading terms. 

The above solution is consistent if the OPE is of the following form 
\bea
\phi_j (x_1, z_1) \phi_0(x_2,z_2) \!\!\! = \!\!E_j^x(x_2,z_2)\ln|x_{12}|+E_j^z(x_2,z_2)\ln|z_{12}| +F_j(x_2,z_2) + ...  \nonumber \\
\phi_j (x_3, z_3) \phi_0(x_4,z_4) \!\!\! = \!\!E_j^x(x_4,z_4)\ln|x_{34}|+E_j^z(x_4,z_4)\ln|z_{34}| +F_j(x_4,z_4) + ... \nonumber \\
\eea

The two point functions of the fields $E^{a}_j$ and $F_j$ are ($j\ne -\half$)
\bea
\label{log}
\la E_j^a E_j^b \ra = 0, \quad 
\la  E_j^z(x,z)F_j(0,0)\ra = \frac{2j+1}{(k-2)}|z|^{-2h}|x|^{-2j}  \\ \nonumber
\la E_j^x(x,z)F_j(0,0)\ra = |z|^{-2h}|x|^{-2j} \\
\la  F_j(x,z)F_j(0,0)\ra = -2|z|^{-2h}|x|^{-2j} \left(\frac{2j+1}{k-2}\ln|z|+\ln|x| \right) \nonumber
\eea

If $j=-\half$ it is consistent at this level to set $E_j^z=0$ and we
have logarithmic blocks in isotopic space only.

Now let us see how logarithmic operators can occur in $AdS_3$ string
theory. The fields $E^{a}_j$ and $F_j$ are degenerate and span the Jordan cell structure of the Lie algebra. This is the well known story in logarithmic CFT's. One has to deal with reducible but indecomposable representation of the corresponding symmetry algebra. So one has to look for extension of $\SL$ modules to include the fields $E^{a}_j$ and $F_j$. 

The $\SL$ module is degenerate for the following values of $j$\cite{Kac:1979}

\be
j_{r,s} = \frac{r-1}{2} + \frac{s-1}{2}(k-2)
\ee  

Null states exist for $r,s \in \CZ$ and one can consistently extend
chiral symmetry algebra to include the fields $E^{a}_j$ and $F_j$
satisfying the correlation functions (\ref{log})\cite{Fjelstad:2002}. The extended module for $j_{1,1}\equiv 0$, is also known as singleton representation. 

Normalizable wave functions in $AdS_3$ exist for $j<0$ only and they
are square integrable only for $j<-\half$. So if one is dealing with
square integrable wave functions, it is sufficient to restrict the
fields with $j<-\half$. However to account for existence of
logarithmic pairs of the fields $E^{a}_j$ and $F_j$ one has to deal
with representation to include fields with $j\le 0$. At $j=0$, one encounters singleton representation. It plays a special role in bulk-boundary correspondence as discussed in next section.

\section{Singletons and logarithmic operators in boundary CFT}

Anti-de-Sitter group $SO(d, 2)$ in $d+1$ spacetime dimension has special representations called singletons. They are special because they saturate the unitarity bound and they have singular flat spacetime limit as discovered by Dirac \cite{Dirac:1963ta}. They can also be considered as topological fields living on the boundary of AdS spacetime \cite{Fronsdal:1974ew,Flato:1980zk}. Since the group $SO(d, 2)$ is the conformal symmetry group in d-dimensions, we can consider them the representations of conformal group in d-dimensions. Singletons are in the indecomposable representation of conformal algebra and  give logarithmic correlation functions on the boundary of $AdS_{d}$\cite{Kogan:1999bn,Ghezelbash:1998rj}).

Theory of singletons can be formulated in terms of Flato-Fronsdal dipole \cite{Flato:1987uh} field which satisfy the following equations of motion,
\bea
(\nabla + m^{2})A + B & = &0 \\ \nonumber
(\nabla + m^{2})B & = & 0.
\eea

These can be derived from the following form of the action in $AdS_{d+1}$
\be
\label{bulkaction}
S = \int d^{d+1}x\sqrt{g}\left(g^{\mu\nu}\p_{\mu}A\p_{\nu}B - m^{2}AB - \frac{\mu^{2}}{2} B^{2}\right)
\ee

with $m^2 = \Delta(\Delta-d)$ and $\mu^2 = (2\Delta-d)/2$. Singleton corresponds to $\Delta = \Delta_0 = \frac{d-2}{2} \Rightarrow \mu^2 = -1$. For $AdS_3, d=2$ and $\Delta = \Delta_0 = 0 = m^2$, so we shall consider the limit $\Delta\rightarrow 0$.

The above form of the action is shown to give logarithmic two point functions on the boundary \cite{Ghezelbash:1998rj} of AdS spacetime. However, it does not make sense to write an action of this form for all $\Delta$ except singletons. So one should work with the following action (with understanding $m^2\rightarrow 0$ for $AdS_3$).
\be
S = \int d^{d+1}x\sqrt{g}\left(g^{\mu\nu}\p_{\mu}A\p_{\nu}B - m^{2}AB + \half B^{2}\right)
\ee

Given the form of singleton action in the bulk of AdS spacetime it remains to determine the pair of fields on the boundary (say C and D) which couple to the dipole fields A and B. Coupling of the fields C and D is of either $\int d^{d-1}x(\alpha A_{0}C + B_{0}D)$ or $\int d^{d-1}x(\alpha A_{0}D + B_{0}C)$.

Choosing the coupling $\int d^{d}x(\alpha A_{0}D + B_{0}C)$ of
boundary operators $C$ and $D$ we get the two point functions of
logarithmic operators $C$ and $D$ on the
boundary \cite{Kogan:1999bn,Ghezelbash:1998rj} via AdS/CFT correspondence. 
\bea
\label{eq:logco}
<C(\vec{x} )C(\vec{x^{\prime}})> & = & 0 \nonumber \\
<C(\vec{x})D(\vec{x^{\prime}})>& = & = \frac{\Delta}{\mid\vec{x}-\vec{x^{\prime}}\mid^{2\Delta}} \\ \nonumber
\alpha^2<D(\vec{x})D(\vec{x^{\prime}})>& = & \frac{\Delta}{2\Delta-d}\frac{1}{\mid\vec{x}-\vec{x^{\prime}}\mid^{2\Delta}}\left[-2ln\left(\frac{\mid\vec{x}-\vec{x^{\prime}}\mid^2}{\epsilon}\right) + \frac{1}{\Delta}\right] 
\eea

Standard normalization of equation (\ref{log}) gives the value of $\alpha = 1/(2\Delta-d)$.

As $(x, \bar{x})$ has the interpretation of boundary co-ordinates, we see that there are logarithmic operators in boundary CFT if they are present in the bulk theory. 

Thus beginning with singleton action in bulk of AdS spacetime, we get
logarithmic correlation functions on the boundary via AdS/CFT
correspondence.

\section{Correlation functions beyond $j = -\half$}

As discussed earlier normalizable wave functions in $AdS_3$ exist for
$j<0$ only and they are square integrable only for $j<-\half$. To
define correlation functions beyond $j = -\half$ one has to take dimension dependent normalization of the fields into account as discussed below.

Consider properly normalized primary fields of the $\SL$
WZNW model

\be
\phi_j(x, z) =  \frac{2j+1}{\pi}\left [(\gamma - x)(\bar{\gamma} - \bar{x})e^{2\phi} + e^{-2\phi}\right ]^{2j}.
\ee

The two point function is completely determined completely by global conformal invariance.

\be
\la \phi_{j_1}(x_1,z_1) \phi_{j_2}(x_2,z_2) \ra = B(j_{1})\delta(j_1-j_2)\mid x_{12}\mid^{4j_1}\mid z_{12}\mid^{-4\Delta_{j_{1}}}
\ee

There is one more $\delta$ - function term on right hand side but it
vanishes if we restrict ourselves to $j<-\half$. Coefficient B(j) is evaluated in \cite{Teschner:1997ft} and is given by
\[
B(j) = -{\pi}^{2j+1}\left(\frac{\Gamma(1-b^2)}{\Gamma(1+b^2)}\right)^{2j+1}\frac{2j+1}{\pi}\frac{\Gamma\left(1+b^2(2j+1)\right)}{\Gamma\left(1-b^2(2j+1)\right)} ;  \                \  b^{2} = \frac{1}{k-2}.
\]

In the supergravity limit $k\rightarrow\infty$ two point function becomes

\be
\la \phi_{j}(x_1,z_1) \phi_{j}(x_2,z_2) \ra = -\frac{2j+1}{\pi}\mid x_{12}\mid^{4j}
\ee

This can be interpreted as two point function of operators on the
boundary which couple to bulk field $\phi_{j}(x, z)$ and is well
defined for all values of $j<-\half$ and vanishes at $j= -\half$. 

Now let us consider two point functions in boundary conformal field theory in the context of $AdS_{3}/CFT_{2}$ correspondence.

Consider the Euclidean action for a massive scaler field in $AdS_{d + 1}$,
\be
S \left[\phi \right] = \half\int_{AdS_{d + 1}} d^{d + 1}x \sqrt{g}\left [g^{\mu\nu}\p_{\mu}\phi\p_{\nu}\phi + m^{2} \phi^{2}\right ] 
\ee

Scalar field in the bulk of AdS spacetime has the boundary limit,
\be
\phi(z,\vec{x}) \rightarrow z^{d - \Delta}\left [\phi_{0}(\vec{x}) + o(z^{2})\right] + z^{\Delta}\left [ A(\vec{x}) + o(z^{2}\right]
\ee

where $\phi_{0}(\vec{x})$ acts as a source function which couples the
background field in the bulk and $A(\vec{x})$ denotes the small
fluctuation around this background and we have denoted boundary points
by a vector and z is the radial co-ordinate in AdS. 

$\Delta$ is given by the roots of the equation
\be
\label{root}
\Delta(\Delta - d) = m^{2}.
\ee

$\phi(z,\vec{x})$ can be constructed from $\phi_{0}(\vec{x})$ 
 via
\be
\phi(z,\vec{x}) = \int d^{d}x^{\prime}K_{\Delta}(z,\vec x, \vec{x}^{\prime})\phi_{0}(\vec{x}^{\prime})
\ee

where $K_{\Delta}(z,\vec x, \vec{x}^{\prime})$ is bulk to boundary propagator \cite{Witten:1998qj,Gubser:1998bc}.
\be
\label{propag}
K_{\Delta}(z,\vec x, \vec{x}^{\prime}) = \pi^{-d/2}\frac{\Gamma(\Delta)}{\Gamma(\Delta - \frac{d}{2})}\left (\frac{z}{z^2+(\vec{x} - \vec{x}^{\prime})^2} \right)^{\Delta},
\ee

The two point function of operators $O_i(\vec{x})$ with conformal dimension $\Delta$, which couple to the fields $\phi_i(z, \vec{x})$ at the boundary is given by
\be
\la O(\vec{x_1}) O(\vec{x_2})  \ra = \frac{1}{\pi^{d/2}}\frac{(2\Delta
  -d)\Gamma(\Delta)}{\Gamma(\Delta -d/2)}\frac{1}{\mid (x_{1}
  - x_2)\mid^{2\Delta}}.
\ee

For $AdS_3$, we have $d=2$ and  $2j = -\Delta$. Equation (\ref{root}) has two roots

\[
\Delta_{\pm} = 1 \pm \sqrt{1 + m^2}
\]

Breitenlohner-Freedman \cite{Breitenlohner:1982jf} stability bound $m^2 > -1$ for a massive scaler field in $AdS_{3}$ implies that $\Delta_{+}>1$ and $\Delta_{-}>0$.

Two point function of operators on the boundary can be defined for $\Delta > 1$ for all $\Delta_{+}$ and is given by \cite{Freedman:1998tz}.

\be
\la O_{\Delta}(x_{1})O_{\Delta}(x_{2})\ra = \frac{1}{\pi}\frac{(2\Delta - 2)\Gamma(\Delta)}{\Gamma(\Delta - 1)}\frac{1}{(x_{12})^{\Delta}}
\ee

The factor $\frac{\Gamma(\Delta)}{\Gamma(\Delta - 1)}$ comes from the normalization of the boundary bulk propagator.
 
To compute two point functions of operators on the boundary CFT which couple to fields
beyond $j=-\half$ barrier, we need to define two point function for the branch $0<\Delta\le 1$, $\Delta =0 $ being the unitary bound, by dimension dependent renormalization of the fields \cite{Klebanov:1999tb}.

\be
\la O_{\Delta}(x_{1})O_{\Delta}(x_{2})\ra = \frac{2}{\pi}\frac{\Gamma(\Delta)}{(\Delta - 1)\Gamma(\Delta - 1)}\frac{1}{(x_{12})^{\Delta}} = \frac{2}{\pi}\frac{1}{(x_{12})^{\Delta}}
\ee

This is correct conformally invariant two point function defined for all $\Delta>0$. 

The fact that two point functions of boundary operators can be defined
for $\Delta<1$ is suggestive that two point functions of $\SL$
primaries can also be defined in the range  $0\ge j \ge -\half$ (Note: $\Delta = -2j$) by defining $\phi_{i}(x, z)\rightarrow \frac{\phi_{i}(x, z)}{2j+1}$. All such values of $j$ violates
Breitenlohner-Freedman stability bound in $AdS_{3}$.

We have considered only real values of $j$ as it gives real conformal
dimensions of the operators in boundary CFT. Meaning of complex $j$
belonging to continuous series is still unclear from the viewpoint of
AdS/CFT correspondence. Thus it is possible to define correlation
functions for all values of $j<0$, which is the correct unitarity bound from the point of view of CFT.

\section{Summary and discussion}

We have discussed the possible occurrence of logarithmic operators in
$AdS_3$ string theory. Origin of logarithmic operators lie in the
fact that some fields are in indecomposable representation of extended symmetry
algebra.

It was noticed that for  $j=-\half$ representation there are no
logarithmic singularities in the coordinate space, but they are present in
isotopic space. It was shown in \cite{Lewis:2000tn} that OPE of $\phi_{-\half}$ with
any other operator doesn't have logarithmic terms and  $\phi_{-\half}$
is analogue of prelogarithmic operator as in Liouville theory. If we
include $\phi_{-\half}$ in the spectrum, we also have to include the
other logarithmic operators in the spectrum, which have logarithmic 
correlation functions. 

The singleton (fields in indecomposable
representation of conformal algebra) action in the bulk gives boundary
two point functions, which are logarithmic in nature. Thus singletons
play a special role in AdS/CFT correspondence.
 
It was possible to define correlation
functions of the operators using a dimension dependent renormalization 
beyond $j=-\half$ representation using AdS/CFT correspondence which are
well behaved for all values of $j$ upto $j=0$. The representation
$j=0$ is special and corresponds to singletons.


\vspace{5mm}

{\bf Acknowledgments:}
We acknowledge the discussion with Ian Kogan and useful correspondence with Alex Lewis. This work is supported by Felix Scholarships and Radhakrishanan Memorial Bequest (University of Oxford, Oxford).


\end{document}